\documentclass[10pt, conference, compsocconf]{IEEEtran}
\ifCLASSINFOpdf
\else
\fi
\usepackage{svg}
\usepackage{amsmath}
\usepackage{amssymb}
\usepackage{booktabs}
\usepackage[normalem]{ulem}
\usepackage{pgfplots}
\pgfplotsset{compat=1.11}
\usepackage{caption}

\usepackage{subfig}
\usepackage{graphicx}
\usepackage[multiple]{footmisc}
\newlength{\bibitemsep}
\newlength{\bibparskip}
\let\oldthebibliography\thebibliography
\renewcommand\thebibliography[1]{%
  \oldthebibliography{#1}%
  \setlength{\parskip}{\bibitemsep}%
  \setlength{\itemsep}{\bibparskip}%
}

\makeatletter
\def\footnoterule{\kern-3\p@
  \hrule \@width 2in \kern 2.6\p@} % the \hrule is .4pt high
\makeatother
\newcommand{\copyrightnotice}[1]{{%
  \renewcommand{\thefootnote}{}% Remove footnote number
  \footnotetext[0]{#1}%
}}

\begin{document}
\title{NPS-AntiClone: Identity Cloning Detection based on\\ Non-Privacy-Sensitive User Profile Data}


\author{\IEEEauthorblockN{Ahmed Alharbi,
Hai Dong, Xun Yi and
Prabath Abeysekara}
\IEEEauthorblockA{School of Computing Technologies, Centre for Cyber Security Research and Innovation, \\
RMIT University, Melbourne, Australia\\
Email: s3633361@student.rmit.edu.au,
hai.dong@rmit.edu.au,
xun.yi@rmit.edu.au, \\
s3693452@student.rmit.edu.au}}


%Telephone: (800) 555--1212, Fax: (888) 555--1212}


\maketitle
\copyrightnotice{Please cite the paper as the following: \textbf{Alharbi, A., Dong, H., Yi, X, Abeysekara, P.: NPS-AntiClone: Identity Cloning Detection based on Non-Privacy-Sensitive User Profile Data. The 2021 IEEE International Conference on Web Services (IEEE ICWS 2021) (2021)} \\
978-1-6654-1681-8/21/\$31.00 ©2021 IEEE \\
DOI 10.1109/ICWS53863.2021.00083}

\begin{abstract}
Social sensing is a paradigm that allows crowdsourcing data from humans and devices. This sensed data (e.g. social network posts) can be hosted in social-sensor clouds (i.e. social networks) and delivered as social-sensor cloud  services (SocSen services). These services can be identified by their providers' social network accounts. Attackers intrude social-sensor clouds by cloning SocSen service providers' user profiles to deceive social-sensor cloud users. 
We propose %\sout{d} 
a novel unsupervised SocSen service provider identity cloning detection approach, \textit{NPS-AntiClone}, to prevent the detrimental outcomes caused by \textbf{such} identity deception. This approach leverages non-privacy-sensitive user profile data gathered from social networks to perform cloned identity detection. It consists of three main components: 1) a %multi-view account generation,
multi-view account representation model, 2) an embedding learning model %a learning the multi-view account representation 
and 3) a prediction model. The multi-view account representation model forms
%generation generates 
three %different views 
different views 
%forming a multi-view that includes 
for a given identity, namely a post view,  a network view and a profile attribute view.
%profile attributes account. The multi-view account includes post view, network view, and profile attributes.
%The generated multi-view then is %combined 
%learned into a single account representation. 
The embedding learning model
%\sout{uses Weighted Generalized Canonical Correlation Analysis which learns a single embedding from the generated multi-view representation}\textbf{
learns a single embedding from the generated multi-view representation using Weighted Generalized Canonical Correlation Analysis. Finally, NPS-AntiClone calculates the cosine similarity between two accounts' embedding to predict whether these two accounts contain a
%form a pair of 
cloned account and its victim. %the account pair. 
We evaluated our proposed approach using a real-world dataset. The results showed that NPS-AntiClone significantly outperforms the existing state-of-the-art identity cloning detection techniques and \textbf{machine learning approaches}.
\end{abstract}


% \fi
\IEEEpeerreviewmaketitle

\section{Introduction}
\textit{Social sensing} is a paradigm that allows crowdsourcing data from multiple \textit{social-sensors} such as humans, %smart phones
smartphones, smart glasses, etc. \cite{rosi2011social}. This sensed data, often termed as \textit{social-sensor data}, is hosted in \textit{social-sensor clouds} (i.e. social networks such as Twitter and Facebook), and can take different shapes \cite{aamir2017social,aamir2017social1}. Examples of social-sensor data include Facebook status messages and Twitter posts. Social-sensor %data 
clouds have, nowadays, become a significant accessible channel to express opinions regarding events and activities \cite{aamir2017social}. Posts may include critical information, particularly images and descriptions regarding incidents or public events \cite{rosi2011social}.

% [Backup]
% %Social-sensor clouds nowadays 
% Nowadays, social-sensor clouds play a very significant role in special events (e.g. sports, crimes, etc.). Social-sensors can publish %thousands even 
% thousands, or even millions of posts (in the form of texts and/or images) over social-sensor clouds. This large amount of information can be abstracted as %social-senor
% social-sensor cloud services (abbreviated as \textbf{SocSen services})  for the convenience of information deployment and reuse \cite{aamir2017social,aamir2017social1}. The aforementioned special events can be described from different perspectives, such as where, when, and what %is happening 
% through the \textit{functional} and \textit{non-functional properties} of SocSen services \cite{9001217}. Meanwhile, social-sensor clouds can also contain a substantial amount of fake and false information sourced from \textit{unreliable} SocSen service providers with malicious intent. %This sort of information 
% Such false information can spread rapidly over social-sensor clouds due to the proliferation of the Internet and %the 
% prevalence of social networks \cite{gupta2013faking,mendoza2010twitter}. %It will
% This can negatively impact thousands of individuals since most individuals tend to believe the information %regarding public events 
% published on social media in regards to public events \cite{aamir2017social1}.

%Social-sensor clouds nowadays 
Nowadays, social-sensor clouds play a very significant role in special events (e.g. sports, crimes, etc.). Social-sensors can publish %thousands even 
thousands, or even millions of posts (in the form of texts and/or images) over social-sensor clouds. This large amount of information can be abstracted as %social-senor
social-sensor cloud services (abbreviated as \textbf{SocSen services})  for the convenience of information deployment and reuse \cite{aamir2017social,aamir2017social1}. The aforementioned special events can be described from different perspectives, such as where, when, and what %is happening 
through the \textit{functional} and \textit{non-functional properties} of SocSen services \cite{9001217}. 
The prevalence of social-sensor clouds has %\sout{also} 
attracted many adversaries attempting to exploit SocSen service providers' identities and deceive users in many ways. \textit{Identity cloning} is one such mode of deception whereby an attacker registers a fake profile of a provider using the provider's identity information to perform malicious activities \cite{bilge2009all}. Examples for such malicious activities include tarnishing the provider's reputation or deceiving the provider's friends to steal sensitive information that is not publicly available \cite{bilge2009all}. For example, it has been reported that the Facebook account of Mark Zuckerberg, Chief Executive Officer of Facebook, has previously been cloned and involved in financial fraud\footnote{https://www.nytimes.com/2018/04/25/technology/fake-mark-zuckerberg-facebook.html}. Another well-known example is that Russian President Vladimir Putin's Twitter account was cloned and attracted over 1 million followers\footnote{https://www.abc.net.au/news/2018-11-29/twitter-suspends-account-impersonating-vladimir-putin/10569064}. In addition, these fake identities can also be used to spread fake and false information sourced from \textit{unreliable} sources rapidly over social-sensor clouds 
%\sout{due to}
accelerated by the proliferation of the Internet and %the 
prevalence of social networks. This can negatively impact thousands of individuals since most individuals tend to believe the information
published on social media in regards to public events \cite{aamir2017social1}.

Identity cloning can take two key shapes in the form of 1) single-site and 2) cross-site identity cloning \cite{bilge2009all}. The former refers to instances where an attacker creates a cloned identity of a SocSen service provider in the same social-sensor cloud. The latter represents instances where an attacker clones the identity of a provider from another cloud. In this work, we primarily focus on \textit{detecting single-site identity cloning}. %\sout{Many identity cloning related crimes have occurred in the past few years. For example, it has been reported that the Facebook account of Mark Zuckerberg, Chief Executive Officer of Facebook, has previously been cloned and involved in financial fraud\textbf{\footnote{https://www.nytimes.com/2018/04/25/technology/fake-mark-zuckerberg-facebook.html}.}}

%Most social-sensor clouds lack automated support for detecting identity cloning. For example, Twitter and Instagram review identity cloning claims upon receiving a valid identity cloning report. However, they have not built an active detection mechanism to detect identity cloning\footnote{https://help.twitter.com/en/rules-and-policies/twitter-impersonation-policy}\footnote{https://help.instagram.com/446663175382270}. Existing research on cloned identity detection mostly relies on both \textit{privacy-sensitive user profile data} and \textit{non-privacy-sensitive user profile data}. Privacy-sensitive user profile data, such as full names, dates of birth, personal photos, etc., cannot be accessed by third party applications via APIs or other means due to privacy restrictions. As a result, the existing methods are inherently less applicable for third party applications to detect cloned identities. In contrast, \textbf{non-privacy-sensitive user profile data}, such as screen names, profile descriptions, etc., is often readily accessible to third party applications, and can be directly obtained using APIs exposed by social-sensor clouds. Assume that an attacker uses a cloned account to log onto a \textbf{third party} website or application. This third party would find it challenging to accurately identify whether or not it is a cloned account using the existing techniques. Hence, there is an urgent need and potential for exploring approaches to detect identity cloning by utilizing only non-privacy-sensitive user profile data. %In addition, 
Most social-sensor clouds lack automated support for detecting identity cloning. For example, Twitter and Instagram review identity cloning claims upon receiving a valid identity cloning report from the end-users. 
%However, they have not built an proactive detection mechanism to detect identity cloning
Currently, these platforms do not have support for automated tools to detect identity cloning\footnote{https://help.twitter.com/en/rules-and-policies/twitter-impersonation-policy}\footnote{https://help.instagram.com/446663175382270}. Existing research on cloned identity detection mostly relies on both privacy-sensitive user profile data and non-privacy-sensitive user profile data. Privacy-sensitive user profile data, such as full name, date of birth, personal photographs of a user, cannot be accessed by third party applications via Application Programming Interfaces (APIs) or other means due to privacy restrictions. As a result, the existing methods \cite{kontaxis2011detecting,devmane2014detection,kamhoua2017preventing} are inherently less applicable for third-party applications to detect cloned identities. Assume that an attacker uses a cloned account to log onto a \textbf{third party} website or application. This third party would find it challenging to accurately identify whether or not it is a cloned account using the existing techniques. In contrast, \textbf{non-privacy-sensitive user profile data}, such as screen name, profile description, etc. of the users, is often readily accessible to third party applications, and can be directly obtained using APIs exposed by social media platforms. Hence, there is an urgent need and potential for exploring approaches to detect identity cloning by utilizing only non-privacy-sensitive user profile data.

Furthermore, most existing identity cloning approaches  employ \textit{simple feature similarity} or \textit{%are 
supervised machine learning%approaches
} to detect cloned identities \cite{kontaxis2011detecting,devmane2014detection,kamhoua2017preventing,goga2015doppelganger}. Simple feature similarity usually employs human-defined metrics to calculate profile attributes' similarities such as TF-IDF based cosine similarity or Jaro–Winkler distance \cite{devmane2014detection,goga2015doppelganger,jin2011towards}. \textit{These metrics are unable to capture the semantics of various literal strings}. These metrics %\sout{core only}\textbf{}
focus only on characters
%\sout{s} %
distance or word frequency. For example, %\sout{these}\textbf{
the aforementioned metrics cannot measure the semantics between the words \textit{king} and \textit{man}, which describe a gender-based relationship. %they 
%these metrics are unable to capture the semantics of various literal strings. %Furthermore, supervised approach requires 
On the other hand, supervised machine learned-based approaches require labelled data samples for 
%\sout{training the classification model}\textbf{}
predictive model training. However, social networks such as Twitter have hundreds of millions of active users. Therefore, \textit{it is time-consuming and needs high labour cost to obtain enough labelled data samples %for identity cloning 
to train machine learning models} \cite{goga2015doppelganger}. 

Different from the existing works, we focus on \textit{learning social network users\footnote{We use the two terminologies -- \textit{SocSen service providers} and \textit{social network users} interchangeably in this article.}' multi-view representations}. Multi-view learning aims to learn a single function to model multiple views and jointly optimize all the functions to improve their generalization performance \cite{zhao2017multi}.
%Multi-view learning aims to learn a single view from multiple views to bring satisfactory learning performance \cite{zhao2017multi}. 
%It is challenging to learn %a user multi-view 
%a multi-view user representation that can comprehensively %reflect the user. 
%resemble the social network account of a user.  
In the context of social network identity, users can be represented by information from multiple views %data 
including users' posts, social relations %networks %(friends and followers) 
(e.g. friends and followers) and users' attributes (e.g. friends count, posts count). The multi-view information comprehensively describes the users and any loss of this information might potentially lead to missing important information \cite{zhao2017multi}. Existing works had only used \textit{part of the multi-view information} discussed above, thereby leading to unsatisfactory performance \cite{devmane2014detection,kamhoua2017preventing}. %employed a part of the above multi-view information. Thus, it might lead to unsatisfactory performance. 
The challenge of learning the %user multi-view
user's multi-view representation lies in the high non-linearity and high-%dimension 
dimensional characteristics of the social media data, %, for example, 
such as users' posts and social relations networks \cite{luo2011cauchy}. Therefore, there exists an urgent need %to propose a better approach 
for alternative approaches %with a stronger feature learning capacity to represent the user multi-view representation.
with a strong ability to learn features that can more accurately represent the multi-view characteristics of the SocSen service providers.
%Existing identity cloning approaches \cite{kontaxis2011detecting,devmane2014detection,kamhoua2017preventing,goga2015doppelganger} employ \textit{simple feature similarity} or \textit{machine learning models}. Traditional feature similarity usually employs human-defined metrics to calculate profile attributes similarities such as TF-IDF based cosine similarity or Jaro–Winkler distance \cite{devmane2014detection,goga2015doppelganger,jin2011towards}, etc. However, they are unable to capture the semantics of various literal strings. Furthermore, machine learning requires a large labeled amount of training data. %and the hyperpa-rameters need to be fine-tuned, which can be a tedious and time-consuming task. 
%Therefore, there is an urgent need to propose a better way with a small effort of feature engineering to represent the profile attributes. 
%Deep learning, as emerging technologies, have demonstrated their overwhelming performance on executing big data processing and analytics tasks in diverse fields \cite{najafabadi2015deep}. Nevertheless, to the best of our knowledge, there has been no application of deep learning in identity cloning detection so far. 

To address the above limitations, we propose a novel unsupervised approach, \textit{NPS-AntiClone}, for SocSen service providers' identity cloning detection. NPS-AntiClone consists of three main components, namely, 1) a \textit{%multi-view account generation
multi-view account representation model}, %account embedding}
 2) an \textit{embedding learning model} and 3) a \textit{prediction model}. %learning 
%learner that learns the multi-view representation of a user account}. 
The %multi-view account generation 
multi-view account representation model generates three different views from the non-privacy-sensitive user profile data of a pair of accounts sharing the same screen name or username. These views can represent a user's comprehensive social network features, which are highly likely to be mimicked by attackers. %These 
The generated views include a \textit{post view}, a \textit{network view}, and a \textit{profile attribute view}. %\textbf{An account can be represented using these views. It is easy for any person to use another person profile photo which is considered as a profile attribute view. However, it is hard to portray another person deep social networks.}
In the post view, we extract %\sout{the}\textbf{a} 
a pre-trained language representation using Sentence-BERT (SBERT),
%\sout{ where \cite{reimers-2019-sentence-bert}. %The 
%A pre-trained language model}\textbf{which} 
which is a model trained on top of a large amount of unannotated data (e.g. a Wikipedia dump). %crop. 
It aims to extract %\sout{the} \textbf{a} %\sout{text}\textbf{
a textual representation 
%\sout{for any specific task} \textbf{
from a given text content and transform it into a form where similarity measures can be applied. For the network view, we %contract 
consider two types of account networks: follower and friend networks, and then we learn the respective network representation\textbf{s} using Node2vec \cite{10.1145/2939672.2939754}. %\textbf{
Meanwhile, %\sout{For}\textbf{for} 
for the profile attribute view, we employ 12 public profile attributes. %The 
The embedding learning model adopts weighted generalized canonical correlation analysis (wGCCA), which learns a single embedding from the generated multi-view representation %NPS-AntiClone then combines these views into a single representation using weighted generalized canonical correlation analysis (wGCCA) \cite{benton2016learning} 
for each account \cite{benton2016learning}. %\textbf{(this part needs to be updated, refer to my change in the abstract)}. 
It then calculates the cosine similarity between the account pair 
%\sout{and predicts}\textbf{to predict}
to predict whether or not the pair of accounts contains a cloned account and its victim based on their similarity. Our proposed approach shows promising performance in the evaluation based on a real-world Twitter dataset.
%The NPS-Profile Builder extracts three categories of features (termed as \textit{non-privacy-sensitive user features}) from the non-privacy-sensitive user profile data of a pair of accounts sharing the same screen name. These include \textit{similarity-based features, general profile-based features}, and \textit{social relation commonality-based features}. It then builds an augmented user profile (termed as an \textit{NPS-Profile}) for each account based on these features and feeds it into a specially designed CNN model to detect whether or not this account is a cloned account. Our proposed approach shows promising performance in the evaluation based on a real-world Twitter dataset. 
Our main contributions can be summarized as follows:
\begin{itemize}
\item[--] We designed a novel unsupervised SocSen service providers' identity cloning detection approach for third party applications/websites, which depends only on %\sout{the features defined from}
non-privacy-sensitive user profile data accessed via social-sensor cloud APIs. 
\item[--] We devised %an account multi-view learning that creates 
an approach to automatically generate a multi-view representation that can comprehensively describe an account from 
%creates a single account representation from 
a post view, a network view, and a profile attribute view.
%We devised a user profile builder that constructs a unique user profile based on a set of features derived from non-privacy-sensitive user profile data. 
\item[--] We adopted a technique that effectively learns a single account
embedding from all the  %combines all the 
multi-view account representations, which is found to be more robust. %\textbf{(this part needs to be updated, refer to my change in the abstract))}.
%We describe a technique to effectively fuse the account multi-view which learn more robust features.
\item[--] We present an exhaustive evaluation carried out atop a real-world dataset we collected from publicly available Twitter data. We also show that NPS-Anticlone outperforms %We collected a real-world dataset and conducted a set of experiments on it. NPS-AntiClone outperforms 
the state-of-the-art identity cloning detection approaches and other machine learning based approaches using non-privacy-sensitive user profile data.
\end{itemize}
The rest of the paper is structured as follows. Section \ref{rw} reviews existing SocSen service studies and state-of-the-art identity cloning detection techniques. Section \ref{rw1} presents the details of our proposed approach. Section \ref{rw2} describes the evaluation process of our proposed approach. Section \ref{rw3} concludes the paper.

\section{Related work}
\label{rw}
%We reviewed 
In this section, we reviewed and analyzed the related work in SocSen services and identity cloning detection. %in this section.
\subsection{Social-Sensor Cloud Services}
%Massive information has 
%Massive volumes of information have been posted over social-sensor clouds (i.e. social networks) via social-sensors (e,g., smart phones with digital cameras), of late. Social-sensor data (e.g. Facebook status and Twitter posts) can record rich event information as a result of ubiquitous social-sensors. The low cost, quick response and broad spatio-temporal coverage of social-senor data make it an effective means for scene reconstruction and analysis \cite{9001217}. However, the applicability of social-senor data is greatly challenged by its data format and source heterogeneity and the huge  amount of available data for processing and management. 
SocSen services employ a service paradigm to abstract social-senor data into small independent functions. Each SocSen service is defined by its functional and non-functional properties. These functional properties include descriptions, tags, etc. embedded in social-sensor data. The non-functional properties contain spatio-temporal and contextual information, as well as qualitative features (e.g. trust, price, etc.) of social-sensor data. SocSen services aim to reduce social-sensor data complexity and also support real-time and efficient access to relevant and high-quality data \cite{aamir2017social}.

SocSen services provide an easy way to access and manage social-sensor cloud data for building the applications of scene analysis. Aamir et al. \cite{aamir2017social,aamir2017social1} proposed a set of SocSen service selection frameworks to select scene-related social media images based on user queries. Social media images are abstracted as SocSen services. These frameworks try to match SocSen services and user queries in terms of their spatio-temporal and contextual correlations. Aamir et al. \cite{9001217,aamir2018social} further proposed a series of SocSen service composition models for scene analysis. These models take into account the spatio-temporal, image metadata and textual features (i.e. comments and descriptions) of the SocSen services to reconstruct tapestry scenes.

\subsection{Identity Cloning Detection Techniques}
%We first reviewed fake identity or spammer detection which are the closest works for identity cloning detection. 
A significant number of approaches had been %proposed 
proposed in the current literature to detect fake identities or spammers on social media \cite{alharbi2021social}. %\cite{alharbi}. 
The most commonly used approaches %to identify fake or spammer accounts is to
%create 
promote creating behavioural profiles to distinguish between trustworthy and untrustworthy users \cite{masood2019spammer,zheng2015detecting}. These behavioural profiles includes characteristics of users such as writing style or following accounts, etc. %Behavioural 
However, behavioural profiles features are not optimal to detect identity cloning, since %an attacker tries 
an attacker trying to clone the identities can mimic their profile attributes. Therefore, we need to employ features that are able to accurately characterize account pairs to detect the cloned identity. Other works use the trust relationship between users in social networks. The main assumption of these techniques is that a fake/spammer account cannot build an arbitrary number of trusted connections with legitimate accounts in social networks \cite{masood2019spammer,al2017sybil}. This assumption might  %do 
not hold true when dealing with %cloning 
cloned accounts. %Cloning accounts try to mimic 
In the context of cloning accounts, attackers attempt to mimic legitimate accounts. Thus, it is much easier for cloned accounts to build connections with the legitimate accounts than other types of fake identities.

A few approaches had been proposed in the current literature to detect identity cloning on social media \cite{alharbi2021social}. Kontaxis et al. \cite{kontaxis2011detecting} proposed a methodology that allows users to see if they have fallen victim to identity cloning. Devmane and Rana \cite{devmane2014detection} designed a methodology to detect identity cloning attacks in both single and cross-site contexts. It searches for similar user profiles and then calculates a similarity index to detect the cloned profiles. Jin et al. \cite{jin2011towards} analysed and characterized the behaviours of the identity cloning attacks. They proposed two profile similarity schemes to detect suspicious profiles. Kamhoua et al. \cite{kamhoua2017preventing} overcame identity cloning attacks by matching user profiles across multiple social media. They used a hybrid string-matching similarity algorithm to find profile similarity. Goga et al. \cite{goga2015doppelganger} proposed an approach that can detect impersonation attacks.  Their approach detects whether a pair of profiles are controlled by the same person or imposter. It compares the impersonation account activity and reputations. It leverages a Support Vector Machine (SVM) binary  classifier  to detect impersonation attacks.

Most of the existing studies detect identity cloning based on both privacy-sensitive and non-privacy-sensitive user profile data. %privacy-sensitive user profile data and non-privacy-sensitive user profile data. 
However, many third party applications or websites %employ social media identities for authentication. They are not able 
that employ social media identities for authentication are not able to access the privacy-sensitive user profile data using social media APIs. Therefore, the performance of these approaches is questionable for these third parties. In addition, most of the existing approaches are built on simple feature similarity models or classic supervised machine learning models. %The proposed approach in this paper, on the contrary, is 
Instead, we propose an unsupervised approach that is able to learn the account multi-view and effectively fuse them to detect identity cloning. %Deep learning technologies have shown their superior performance on processing and analyzing big data in many application domains \cite{najafabadi2015deep}.  

\section{Proposed Solution}
\label{rw1}
In this section, we present a detailed overview of the proposed NPS-AntiClone approach and its key components. 
\subsection{Overview}
\begin{figure}[]
  \centering
  \includegraphics[width=\columnwidth]{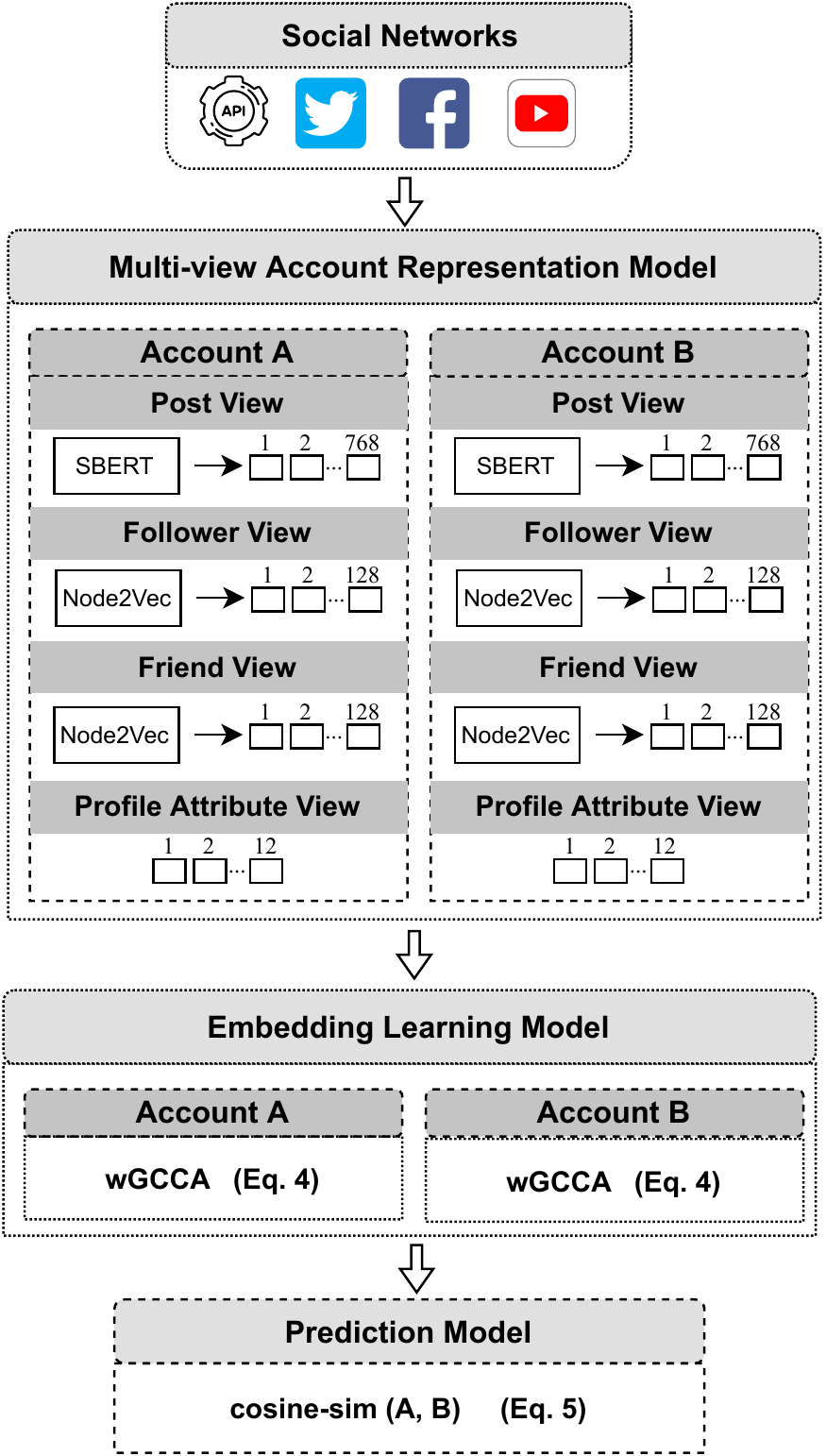}
  \caption{The workflow of NPS-AntiClone}
  \label{over}
\end{figure}

Given a set of social media accounts, we aim to detect identity cloning using multi-view representation
%\textbf{%\sout{data}
 of %the 
an account. Firstly, we generate different %views: posts, network, and profile attributes
views based on 1) posts, 2) network and 3) profile attribute of a given social media account. For the post view, we use Sentence-BERT (SBERT) to extract the account's post representation. For the network view, we build two networks, namely, followers and friends and then extract the network representation using Node2vec, which is an algorithmic framework for graphic representation learning. For the profile attribute view, we extract 12 public profile attributes from the account public profile. %\textbf{We then combine these views into a single representation using weighted GCCA (wGCCA), which learns a single representation from multiple views}. 
We then use wGCCA to learn a single embedding from all these views. We then calculate the cosine similarity between an account pair to predict whether or not it contains an account and its replica. Figure \ref{over} shows the workflow of NPS-AntiClone. %each account pair are an account and its replica.
\subsubsection*{\textbf{Suspicious Account Pair Generation}}
%In a real-world scenario, social media has a very large number of users. 
Social media networks typically contain a very large number of users. Therefore, we need to generate suspicious account pairs for identity cloning detection. %Following the existing work \cite{goga2015doppelganger,perito2011unique}, we search for account pair who has a similar username or screen name to each other. 
Here we adopted a mechanism commonly used in existing works \cite{goga2015doppelganger,perito2011unique}, whereby we search for a pair of accounts that contain a similar username or screen name. For each suspicious account pair, we generate %the 
multi-view account representation.
%\textbf{%not data }. %A matched two accounts only mean that they have a potential likelihood to 
If a match is found between the two accounts in an account pair, it implies that the two accounts have a potential likelihood to %be the same individual 
represent the same individual but their true labels are unknown.
\subsection{%Multi-View Account Generation
Multi-view Account Representation Model}
Our goal is to generate %the 
a multi-view account representation for a given user account by combining several views that %corresponding 
correspond to the %account non-privacy-sensitive profile data. 
non-privacy-sensitive profile data of the account. We use three different %views: 1) account's posts', 2) account's network and 4) account's attributes. 
views based on the account's 1) post view, 2) network view and 3) profile attribute view. 
These views can comprehensively represent a user account, which are highly likely to be mimicked by attackers.
%An attacker can mimic a single view such as the profile public attribute. For instance, the attacker can use the account's profile photo to deceive social network users. However, it is hard to mimic an account's deep social network relationship. For example, the attacker can follow any account that the original account follows but it is hard for the attacker to mimic the deep social interactions for the account's network (e.g. friends of the account's friends).
Then, wGCCA is used to learn a single embedding from these views \cite{hotelling1992relations}.  %\textbf{Then, these views are combined using Canonical Correlation Analysis (CCA) \cite{hotelling1992relations} to obtain a single representation}. 
We describe each view in detail in the following subsections.
\subsubsection{Post View}
To construct the post view, we extract %the 
a pre-trained language representation for each account. In this paper, we use the Sentence-BERT (SBERT) \cite{reimers-2019-sentence-bert} to generate the vector-space representations of the account posts. SBERT is an adjustment of the pre-trained bidirectional encoder representations from transformers network (BERT) \cite{devlin2018bert}. BERT is a technique for pre-training language representations. It can be used for extracting high-quality language features or can fine-tune these models on a downstream task such as classification, question answering, etc. \cite{devlin2018bert}. These pre-trained models are very efficient in extracting the text representation for any specific task from the transformer architecture \cite{qiu2020pre}. SBERT uses siamese and triplet network structures to obtain semantically meaningful sentence representation. %SBERT models  
Models trained using SBERT are also based on transformer networks similar to the BERT models. In addition, SBERT also adds a pooling operation to the output of BERT to obtain a fixed-sized sentence representation. The sentence representation is usually obtained by computing the mean of all output vectors. %We gathered all the account's posts as $T = (t_1, ..., t_n)$ of $n$ posts to obtain 
We gathered $n$  posts publicly available in a given account, denoted as $T = (t_1, ..., t_n)$ to obtain the posts' representation for a user $u$. The representation of %each $t$ is obtained 
each post $t_{i} $($i \in 1,..,n$) is obtained from the pre-trained language representation. 

\begin{figure}[]
  \centering
  \includegraphics[width=0.9\columnwidth]{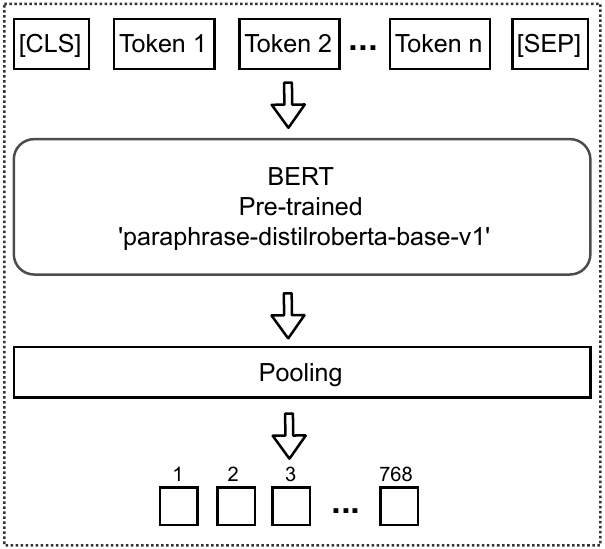}
  \caption{The architecture of SBERT}
  \label{bert_figure}
\end{figure}

Figure \ref{bert_figure} shows the architecture of SBERT. Each %account post $t$ 
post $t_{i}$ is %tokenized 
first tokenized into a single word $w_{i}$  %$w$ 
and then a special token named [CLS] is added to mark the beginning of a sentence. %[CLS] is added at the first position and 
[SEP] is also a special token added to mark the end of the sentence. Then, %set of tokenization 
a set of tokenized words is passed through BERT to embed fixed-sized sentences. Then, in the pooling layer, a mean aggregation (see Equation \ref{bert_mean}) is applied to generate $t$ representations. The mean aggregation %was proved 
is known to have higher performance compared to CLS aggregation or max \cite{reimers-2019-sentence-bert}. The output of each %account 
post is made up of 768 dimensions, which are the default setting of BERT.
%(uncased) has a default 768 hidden layer \cite{devlin2018bert}. Therefore, we choose 768 as the posts dimension size.}. 
Finally, for each %account's posts 
post in an account, we %finally 
compute the mean of all the posts' representation $T$  to get  an aggregated representation, as below. %the final posts representation.
% \begin{equation}\label{bert_mean}
%     \begin{split}
%     t = \frac{1}{|T|}\sum_{w \in T} w
%     \end{split}
% \end{equation}
\begin{equation}\label{bert_mean}
    \begin{split}
    t = \frac{1}{|T|}\sum_{t_{i} \in T} w_{i}
    \end{split}
\end{equation}
\subsubsection{Network View}
%An account network 
A network of accounts is a set of users that %have 
engage in various kinds of interactions, such as friendship, retweet, etc. within %the 
a %social network
social network, which can be presented as a networked graph. %For example, 
If an %account 
account in this network %has an interaction
interacts with another account, there will be an edge between them in the %network
corresponding networked graph. We %contract 
consider two types of account networks: 1) \textbf{follower network:} In this network, if an account follows a specific user such as a celebrity or a friend, there will be an edge between them, and 2) \textbf{friend network:} In this network, if an account gets a follow by another account, there will be an edge between them. Inspired by the success of the graph representation, we utilize the Node2Vec %\cite{10.1145/2939672.2939754} 
to learn the %account networks representation
network representation, or in other words, network view of an account \cite{10.1145/2939672.2939754}. Node2vec is a broadly applied unsupervised representational learning approach for graphs. It tries to maximize the log-probability between %nodes neighbours 
neighbours of a node, or in other words, accounts with an edge between them, via a biased random walk method. %The 
%Node2vec will generate sequences of nodes (directed subgraphs) like sentences are feed into a skip-gram model  to obtain nodes feature representation \cite{DBLP:journals/corr/abs-1301-3781}
Node2vec will generate sentences from the graph. A sentence is a list of nodes (directed subgraphs) that makes a corpus. Then, these corpora are feed into a skip-gram model to obtain node feature representation \cite{DBLP:journals/corr/abs-1301-3781}.
%}\textbf{(not a sentence)}.
Skip-gram is an unsupervised learning approach that is applied to obtain the most related words for a given word. It is also  applied to predict the context word for a given target word.
%We feed these two networks into Node2Vec \cite{10.1145/2939672.2939754} to learn the account networks embeddings. Node2vec is a widely applied unsupervised representational learning approach for graphs. It maintains proximity between nodes via a biased random walk method and inputs the generated node sequences into the skip-gram model \cite{DBLP:journals/corr/abs-1301-3781} to obtain nodes embedding.

%Consider a given an account's follower network $G(V,E)$
Let us denote an account's follower network as $G(V,E)$, where $V = [v_1,v_2,...,v_n]$ represents a set of nodes (i.e. users), $n$ represents the total number of nodes and $E$ represents a set of edges (i.e. social interactions). For every node $v \in V$, let $Nu(v) \subset V$ represent a network neighborhood of node $v$ that is constructed over a neighbourhood sampling strategy $S$. $S$ is a flexible biased random walk technique that can trade off between local and global views of the network in a depth-first (DFS) as well as breadth-first (BFS) fashion. Node2Vec can learn the node's representation by optimizing the objective function as follows:
\begin{equation}\label{nod2vec}
    \begin{split}
    \max_{f} \sum_{v \in V} log Pr(N_S(v)|f(v))
    \end{split}
\end{equation}

where %$f: V \rightarrow \mathbb{R}^d$ 
\( f: V \rightarrow \mathbb{R}^d \) represents a mapping function that aims to learn mapping nodes to feature representation %($V \in \mathbb{R}^d$)
(\( V \in \mathbb{R}^d \)). Node2Vec uses stochastic gradient descent (SGD) as its underlying optimization method to learn the node representation mapping function $f(·)$.
%We can learn the node embedding mapping function $f(·)$ by utilising stochastic gradient descent (SGD) for optimizing this objective. 
We utilize Node2vec to generate the follower network and friend network of the account since Node2vec can learn the high-level feature representations %of a node. These generated features reflect the 
that include deep social interaction of accounts. This information can be helpful for differentiating between an account pair. We assume that it is hard for an attacker to mimic these deep social interactions. 
\subsubsection{Profile Attribute View}
\begin{figure*}[htbp]
  \centering
  \subfloat[post view]{\includegraphics[width=0.5\columnwidth]{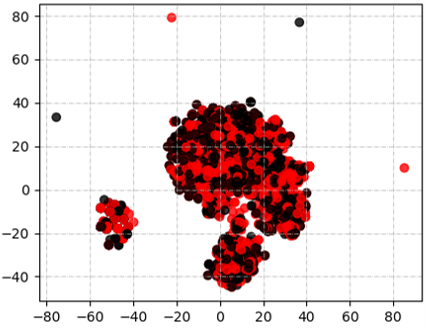}\label{fig:f1}}
  \subfloat[friend network View]{\includegraphics[width=0.5\columnwidth]{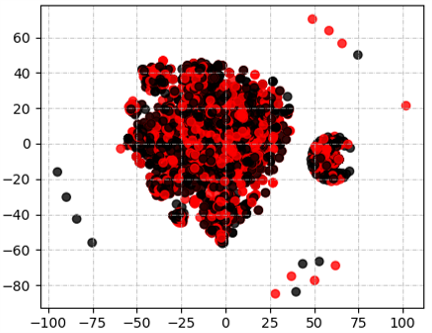}\label{fig:f2}}
  \subfloat[follower network View]{\includegraphics[width=0.5\columnwidth]{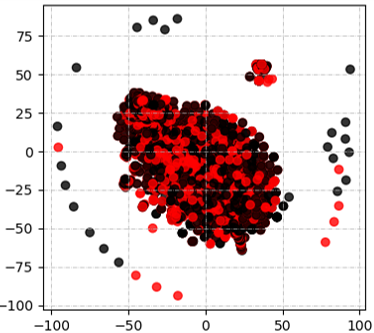}\label{fig:f1}}
  \caption{t-SNE visualization of the accounts representations created using a) post view representation, b) friend network representation and c) follower network representation. %Cloned accounts are the red and the non-cloned accounts are the black
  Cloned accounts are depicted in red and non-cloned accounts are depicted in black.}
  \label{tsne}
\end{figure*}

%We extract 
To construct the profile attribute view of an account, we extract 12 public profile attributes of %the 
an account to create  an attribute vector. These attributes can categorize the activities and reputation of an account. For example, the number of tweets can indicate %that the account activity 
the activities of an account, while the number of followers can indicate the %account reputation. 
reputation of an account. We consider the following profile attributes:
\begin{enumerate}
    \item Friend (following) count: The number of users that the account follows.
    \item Follower count: The number of followers that the account has.
    \item Favorite count: The number of tweets that the account has liked. %The count of favorites tweets that the account has liked.
    \item Tweet count: The number of tweets (including retweets) that the account has posted.
    \item List count: The number of public lists of which the account is a member.
    \item Account age: The %account age 
    life time of the account to-date in months from the date of the account registration.
    \item Profile background: A binary value that shows whether the account has not altered the background or the theme of their profile.
    \item Profile image: A binary value that shows whether the %the 
    account has not uploaded their profile image and that a default image is used.
    \item Has profile description: A binary value that shows whether the account has added a description to their profile.
    \item Profile URL: A binary value that shows whether the %the 
    account has added a URL to their profile.
    \item Screen name length: The %screen name length 
    length of the screen name of the account.
    \item Description length: The %description length 
    length of the description of the account.
\end{enumerate}
Figure \ref{tsne} shows the visualization of all proposed views using t-Distributed Stochastic Neighbor Embedding (t-SNE) \cite{van2008visualizing} on each view vector. t-SNE is an unsupervised, non-linear approach %fundamentally 
predominantly utilised to ease data visualization and reduce dimensionality. It %shows the intuition 
provides an intuitive view of how the data is arranged in a high-dimensional space. Cloned accounts and non-cloned accounts are represented by the red and black circles, respectively. It can be inferred that each of the proposed views can be employed to calculate the profile similarity between the accounts and that %if we can effectively combine 
 we can get a better profile similarity performance by effectively aggregating the proposed views. 
\subsection{%Learning the Multi-View Account
Embedding Learning Model}
The %described views in the previous Section 
views described in the previous section includes some knowledge that might be utilised to detect cloned accounts. %Though
Employing a single view might result in the loss of valuable information. %which helps 
%which would otherwise help enhance the detection performance. 
A simple and naive technique is to %concatenate all the views together
consolidate all these views by concatenating them together. %However, a drawback of this technique is that it will generate a very large account representation. It  might also disregard the complementary existence of the knowledge included %in one of the views, particularly the account network views
%in some views such as the network view of an account. 
However,  a drawback of this concatenation is that it causes over-fitting on a small training sample because of the large account representation. It might also disregard the meaningful knowledge included in the views because each view has a specific statistical property \cite{xu2013survey}.
Therefore, we use an alternate technique %to combine the proposed views into a comprehensive single representation using CCA
named CCA to %combine 
learn a comprehensive single embedding from the proposed views \cite{hotelling1992relations}.
%\textbf{learn the proposed views into a comprehensive single representation \cite{hotelling1992relations}}. 
CCA finds maximal information from two sets of views and generates an %combined
%representation
embedding.
CCA has been applied to several multi-view data learning problems and achieved successes \cite{xu2013survey}.
The limitation of CCA is that it can only %combines 
learn two views. %we employ Generalized CCA (GCCA) which learns a single %representation
%embedding from multiple views  \cite{carroll1968generalization}. %which is utilised to analyze various views simultaneously.
Therefore, we employ Generalized CCA (GCCA) that allows multi-view based learning. GCCA contains many variations, such as
%that proposed by Carroll. Various GCCA models have been proposed 
\cite{robinson1973generalized,tenenhaus2011regularized,carroll1968generalization}.
Here we employ Carroll \cite{carroll1968generalization}'s GCCA, since it is based on an eigenequation that is computationally straightforward and efficient. 
% It is also almost similar to the familiar multivariate techniques such as principal component analysis (PCA).
The GCCA objective can be formulated as follows:
\begin{equation}
\label{gcca}
    arg \min_{G_i,U_i} \sum_i \parallel G-X_iU_i \parallel_F^2 \qquad s.t. G' G = I 
\end{equation}
where $X_i \in \mathbb{R}^{\ n\times d_i}$  corresponds to the data matrix %for 
of the $i^{th}$ view, $G \in \mathbb{R}^{\ n\times k}$ contains all learned account embedding 
%\textbf{%representation embedding} 
and $U_i \in \mathbb{R}^{\ d_i\times k}$ maps from the latent space to the observed view $i$. However, each view can be more or less important for detecting identity cloning. Therefore, we use weighted GCCA (wGCCA) \cite{benton2016learning}. wGCCA adds weight $w_i$ for each view $i$ to Equation \ref{gcca} as follows:
\begin{equation}
\label{wgcca}
    arg \min_{G_i,U_i} w_i\sum_i \parallel G-X_iU_i \parallel_F^2 \; s.t. G' G = I, w_i\geq0 
\end{equation}
where $w_i$ %represents the view's importance of the $i^{th}$ view. 
represent the weight of a view and this weight shows the view's importance. The columns of $G$ are the eigenvectors of $\sum_i w_i X_i({X_i}'X_i)^{-1}{X_i}'$ and the solution for $U_i=({X_i}'X_i)^{-1}{X_i}'G$.
\subsection{%Predication 
Prediction Model}
We used the cosine similarity to compute the similarity between the account pair as follows:
\begin{equation}\label{cosine}
        \begin{split}
   \cos(\bf{A_1}, \bf{A_2}) = \frac{\bf A_1\cdot\bf{A_2}}{||{\bf A_1}||\cdot||{\bf A_2}||}
   \end{split}
\end{equation}
where \(\bf A_1\) and \(\bf A_2\) refer to the account pair %representation.
embedding. A threshold will be selected to determine whether a given account pair contains a cloned account and its victim based on their similarity score.

\section{Evaluation}
\label{rw2}
A set of experiments was performed to evaluate and analyse the effectiveness of NPS-AntiClone against existing state-of-the-art identity cloning detection approaches. In addition, we also evaluated several candidate machine learning models to assess their performance in this particular problem context. %and justify the use of deep forest as the cloned identity predictor of NPS-AntiClone. 

All the experiments were conducted on a computer with Intel Core i5 1.80 GHz CPU and 16 GB RAM. In addition, all the candidate models compared were implemented in Python. We used the SBERT package\footnote{https://github.com/UKPLab/sentence-transformers} to extract the pre-trained language  %representation 
representations used by NPS-AntiClone. We also used the StellarGraph package\footnote{https://github.com/stellargraph/stellargraph} to extract the Node2vec %representation 
representations used by NPS-AntiClone. We used Tensorflow\footnote{https://www.tensorflow.org/} (v1.14.0) library written in Python to implement the %other 
deep learning models evaluated and scikit-learn\footnote{https://scikit-learn.org/stable/} (v0.22.1) for the other machine learning models evaluated.
All the machine learning models were run for 10 rounds with different random permutations of the data. The results were presented as an average computed across %the all rounds of experiments
all rounds of the experiments, together with standard deviation.

\subsection{Dataset}
To the best of our knowledge, there %is 
are no readily and publicly available datasets that could be used for evaluating identity cloning in the domain of social networks. Most existing works, albeit limited, evaluated their proposed approaches on simulated data. Therefore, to evaluate NPS-AntiClone, we developed a dataset via %authroized 
authorized non-privacy sensitive user profile data fetched from Twitter APIs\footnote{https://developer.twitter.com/en/docs}. We collected 4,030 public Twitter accounts (2,015 cloned accounts and their corresponding victim accounts) from \footnote{https://impersonation.mpi-sws.org/}. Table \ref{data_set} summarizes the detailed statistics. We also randomly collected 20,152 non-cloned public Twitter accounts to add noises to the dataset. 
%We take into consideration the creation of the account pair. 
In total, we have 24,182 public Twitter accounts. The dataset was randomly split with a 80:20 training-to-split ratio in order to derive training and test datasets for all the supervised machine learning models.

\begin{table}[]
\centering
\caption{Statistics of the dataset}
\label{data_set}
\resizebox{\columnwidth}{!}{%
\begin{tabular}{p{0.09\columnwidth}p{0.12\columnwidth}p{0.1\columnwidth}p{0.2\columnwidth}p{0.18\columnwidth}}
\toprule
\textbf{Dataset} & \textbf{\#Accounts} & \textbf{\#Tweets} & \textbf{\#Friend network} & \textbf{\#Follower network} \\ \toprule
Twitter & 4030 & 948,144 & 514,534 & 365,687 \\\toprule
\end{tabular}%
}
\end{table}

%\subsection{Evaluation Metrics}
%\textbf{The aim of the paragraph was to show which cosine similarity score our model uses and we choose 0.1 as it achieves the optimal performance results. }Our aim is to calculate the profile similarity for the account pair. %When calculating the cosine similarity between the account pair, we choose the cosine similarity between the account pair. 
%\sout{When the similarity score of an account pair is greater than 0.8, we label them as positive and negative otherwise. We chose the similarity score based on the assumption that a cloned account will try to clone an honest account by 80\% or more. In addition, we have evaluated all the cosine similarity scores in the following subsection. We compute the Precision, Recall and F-score against all the similarity scores. Hence, we chose 0.8 as the similarity score. } \textbf{We have evaluated all the cosine similarity scores in the following subsection. We have decided the optimal cosine similarity score threshold by computing the Precision, Recall and F-score against all the similarity scores. We found that 0.1 is the optimal cosine similarity score. Hence, we chose 0.1 as the similarity score.}
%We select the account pairs only if the cosine similarity score of the screen names or usernames of the two corresponding accounts is over 80\%. We assumed that a cloned account is more likely to share the same screen name or username with the original account. We have 3652 positive candidate pairs and 512 negatives candidate pairs. 

\subsection{Other Approaches Evaluated}
We compared and evaluated NPS-AntiClone %with 
against the following approaches %, including 
which include a set of state-of-the-art identity cloning detection approaches, supervised machine learning approaches and variants of NPS-AntiClone:
\begin{itemize}
\item \textbf{Basic Profile Similarity (BPS) \cite{jin2011towards}:} This technique calculates the number of similar public attributes and %\sout{the number of} 
    common friends between a given user profile and its %\sout{the}
    suspected cloned account(s).
    %\sout{pair}. 
   Our experiments  %\sout{W}
    only used the number of common friends in the friend list in the compared accounts%,
    since the recommended and excluded friend lists are privacy-sensitive user profile %data
    that is not available publicly.
    %\sout{because}
    % \textbf{as} we were not able to collect the recommended %\sout{friend list}
    % and excluded friend list\textbf{s} for %\sout{each}
    % \textbf{every} account. 
    Further, %\sout{I}
    in our dataset, each account has 13 public attributes. We also set \(\mu = 0.0154\), \(\varepsilon = 13\), and \(\lambda = 0.03\).
    % \textbf{ If the profile similarity of account pair less than $\mu$, the account pair will be considered not cloned profile. $\mu$ is calculated based on the smallest number of the attribute similarity of account pair $a$ and the smallest number of the common friends of account pair $b$ (see Equation \ref{eq1234}). $\varepsilon$ is the number of public attributes (raw data) in our dataset.  $\lambda$ is calculated based on the smallest $a$ and largest $b$ number of common friend of the account pair (see Equation \ref{eq1234}).}
    If the profile similarity of a given pair of accounts is less than $\mu$, then they are considered as non-cloned. For a given pair of accounts, $\mu$ is calculated based on the smallest number of the attribute similarity $a$ of the account pair and the smallest number of common friends $b$ of the account pair (see Equation-\ref{eq1234}). $\varepsilon$ is the number of public attributes (raw data) in our dataset.  $\lambda$ is calculated based on the smallest $a$ and largest $b$ number of common friends in the account pair, respectively (see Equation-\ref{eq1234}).
    \begin{equation}\label{eq1234}
    \mu,\lambda = \frac{(a)^2+(b)^2}{\sqrt{k^2+x^2}}
\end{equation}
where $k$ and $x$ are adjusted based on the contribution of similar public attributes and the number of common friends.  We assigned 0.5 for $k$ and $x$ to make them contribute equally.
    \item \textbf{Devmane and Rana \cite{devmane2014detection}:} This technique compares names, education, profile photos, places lived,  birth date, workplace, gender, photos added to the profile, number of friends/connections. We only compared the screen name, places lived (location) and the number of friends/connections that can be fetched using Twitter APIs. We then calculated a similarity index for a given pair of accounts. The original work does not show adequate details on the type of similarity index they used in their proposed approach. Therefore, we adopted the screen name and location similarity technique used in our work to calculate the similarity index %in 
    of the compared user profiles.
    \item \textbf{Goga et al. \cite{goga2015doppelganger}:} This technique compares the time at which the pair of accounts were created and the reputations of the accounts in terms of popularity and social influence, and detects whether an account pair has an impersonated account. To determine this, it uses four different types of features, namely,  profile similarity, social neighbourhood overlap, time overlap accounts and differences between accounts. It then trains an SVM classifier, with a linear kernel to classify if a given account is impersonated. In our experiments, we used all the  recommended features in the original work to train the SVM classifier. 
    \item \textbf{Kamhoua et al. \cite{kamhoua2017preventing}:} This technique compares friends list similarity and calculates attribute similarity using a %modified 
    similarity metric called Fuzzy-Sim. To calculate the attribute similarity, it considers the following attributes: name, education, city, age, workplace, gender, and friend list. In our experiments, however,  we only compared the screen name, city (location), and the friend list provisioned by Twitter APIs. We then calculated the attribute  similarities and friend list similarity using FuzzySim  with the same threshold values recommended by the original work (0.565 and 0.575). 
    \item \textbf{Zheng et al. \cite{zheng2015detecting}:} This is a typical spammer detection model. It employs 18 features, which are a combination of profile-based features such as the number of followers, the number of followers, and content-based features such as the average number of hashtags. It then trains an SVM classifier, with a Radial Basis Function (RBF) kernel to %classify spammer and non-spammer
    classify an account as a spammer or non-spammer account. In our experiments, we used all the recommended features recommended in the original work to train the SVM classifier.
    \item \textbf{Other Supervised Machine Learning Models:} We use the following supervised machine learning and deep learning models in order to predict whether %the 
    a given account pair form a cloned and non-cloned pair. These models are Adaboost (ADA), Convolutional Neural Network (CNN), Deep Neural network (DNN), K nearest neighbours (KNN), Logistic Regression (LR), Multi-layer Perceptron (MLP), Random forest (RF) and SVM with a linear kernel. Instead of calculating the cosine similarity between the account pair using Equation \ref{cosine}, we feed the classifier with the wGCCA output to predict whether the account pair %form
    consists of a cloned account and its victim.
    \item \textbf{NPS-AntiClone$_{concat}$:} %We use an alternative way to combine all the views. We concatenate all the views into a single account representation of 
    In this approach, we concatenate all the views in order to combine them into a single account representation of the size $n=1,036$. 
    \item \textbf{NPS-AntiClone$_{posts}$:} %is a variant of NPS-AntiClone. It only employs the posts representation.
    This approach is a variant of NPS-AntiClone that only uses the post view of the accounts to detect an account pair with a cloned account and its victim.
    \item \textbf{NPS-AntiClone$_{net(F,FL)}$:} %is also a variant of NPS-AntiClone. It only employs 
    This is also a variant of NPS-AntiClone that only employs the network's view representation. Additionally, we %try 
    attempt to use the follower network representation and friend network representation separately, and denote them as NPS-AntiClone$_{FL}$ and NPS-AntiClone$_{F}$ respectively. 
    \item \textbf{NPS-AntiClone$_{PA}$:} This is a variant of NPS-AntiClone that only employs the profile attribute view of the accounts.%. It only employs the extracted profile attributes.  
\end{itemize}
 
\subsection{%Parameter Setup
Hyperparameter Tuning}
%All the supervised machine learning models' hyper-parameters have been properly tuned to achieve their optimal performance. 
All the hyperparameters of the supervised machine learning models were properly tuned to obtain their optimal performance. %The dataset was randomly split with a 80:20 training-to-split ratio in order to derive training and test datasets for all the supervised machine learning models. 
Table \ref{parm} shows the hyperparameter values used for the machine learning and deep learning algorithms evaluated. 
We set the parameters of the %state of the art 
state-of-the-art approaches following the original works. 

For the proposed NPS-AntiClone, 
 %we assumed that a cloned account is more likely to share the same screen name or username with the original account. Therefore,
the suspicious account pairs were selected only if the similarity score of the screen names or usernames between a pair of accounts is over 80\%, by following the existing work \cite{goga2015doppelganger}. %Finally, we have 3,652 positive account pairs and 512 account candidate pairs, where a positive pair account refers to a pair of cloned accounts and its victim account while a negative account pair means otherwise.
We used `paraphrase-distilroberta-base-v1'\footnote{https://huggingface.co/sentence-transformers/paraphrase-distilroberta-base-v1} as the pre-trained model for SBERT. This pre-trained model was trained on millions of paraphrase sentences. The dimension size of the post representation using SBERT is $768$, which is a default setting of the model. The default dimension size of the Node2vec for both the follower network and friend network is $128$. We also use the probability for moving away from source node $q = 2$, the probability of returning to source node $p = 0.5$, the number of random walks per root node $n= 10$ and the maximum length of a random walk is $15$. All the profile attribute views were normalized to range 0 and 1. The weight $w$ of the wGCCA is set as $[0.25, 0.5, 0.5, 0.25]$. The optimal threshold of a cosine similarity score between two account embedding is 0.1 according to our experiment (described in Section. IV.D) 
\begin{table}[]
\centering
\caption{Hyperparameter values used for the candidate machine learning and DL algorithms}
\label{parm}
\resizebox{\columnwidth}{!}{%
\begin{tabular}{p{0.1\columnwidth}p{0.7\columnwidth}}
\toprule
\textbf{Model} & \textbf{Parameter} \\ \toprule
ADA & estimators = 50 \\
CNN & 10 layers, filters = 64, kernel size = 2, pool size = 2 \\
DNN & 5 layers (250, 200, 50, 1)\\
KNN & neighbors = 5 \\ 
MLP & solver = adam, activation = relu \\
RF & estimators = 50 \\
SVM & kernel = linear \\ 
\toprule
\end{tabular}%
}
\end{table}

\subsection{Results and Discussion}
We evaluated and compared the performance of all the aforementioned models using Precision, Recall, F1-Score and F2-Score as our Key Performance Indicators (KPIs). Precision measures the percentage of predicted account pairs (a cloned account and its victim) that are correctly detected. Recall measures the percentage of the true account pairs that are correctly detected. 
F1-score is the harmonic mean of precision and recall. F2-score puts more emphasis on identifying as many account pairs with a cloned account in the dataset as possible. 
\subsubsection{Overall performance} 
The results of %our 
the experiments showed that our proposed approach clearly outperforms the existing state-of-the-art identity detection techniques across all the KPIs evaluated (see Table \ref{zzz}). Based on the optimal threshold (i.e. 0.1), our proposed approach achieved a Precision of 88.70\%, Recall of 82.83\%, F1-Score of 85.66\%, and F2-Score of 83.94\%
%\textbf{(are they based on the same similarity threshold? if the thresholds are different, you need to provide the precision, recall, f-1 and f-2 in different thresholds ) Our model uses a 0.1 similarity score and all results based on this similarity score}
, respectively, which were 13.58\%, 10.29\%, 11.86\% and 10.9\% higher than that of the BPS \cite{jin2011towards}, the second best-performing approach. In addition, our proposed approach %outperforms
also outperformed the other existing models in %the
a recall-prioritized situation. A recall-prioritized scenario where a more emphasis is put on identifying as many account pairs with a cloned account in the dataset as possible. The identity detection approaches proposed by Kamhoua et al. \cite{kamhoua2017preventing} and Devmane and Rana \cite{devmane2014detection} performed poorly because they only focus on %the simple profile attributes similarity 
a simple technique to compute profile attribute similarity and ignores the network information and  contents of the account. %the account's content. %use lower numbers of non-privacy-sensitive user features. 
BSP \cite{jin2011towards}, on the other hand, depends on the simple profile attributes-based similarity and common friends between the account pair, but it does not consider the effect of the network information and the account's content. Goga et al. \cite{goga2015doppelganger} depends on a traditional similarity technique, which is based on human defined metrics, to compare between two accounts. Zheng et al.'s model \cite{zheng2015detecting} only focuses on spammer detection. It detects whether an account is a spammer or not by employing a set of features that compares spammer behaviour patterns. Likely, the behaviour of cloned identity patterns differs from spammer behaviour. Cloned identity tries to emulate the legitimate account behaviour %thus Cloned identity is hard to detect
and therefore, is hard to detect by the spammer detection techniques. NPS-AntiClone is an unsupervised approach that employs non-privacy-sensitive account multi-view representation. %data.
The account multi-view representation %data 
can capture both the semantic and literal characteristics of the account, which can be attributed to its superior performance compared to the spammer detection model proposed by Zheng et al.    %For the supervised  spammer detection  model \cite{zheng2015detecting}, it only employs a simple set of features. NPS-AntiClone uses a multi-view embedding, which can capture the differences between the account pair. %The obtained results indicate that NPS-AntiClone is more adaptable to the context that all the user profile data employed for identity cloning detection is the non-privacy-sensitive user profile data provisioned by social-sensor cloud APIs.
\begin{table}[]
\centering
\caption{
Comparison with state of the art identity cloning approaches. }
\label{zzz}
\resizebox{\columnwidth}{!}{%
\begin{tabular}{p{0.31\columnwidth}p{0.09\columnwidth}p{0.09\columnwidth}p{0.12\columnwidth}p{0.12\columnwidth}}
\toprule
\textbf{Model} &   \textbf{Precision (\%)} & \textbf{Recall (\%)} & \textbf{F1-Score (\%)} & \textbf{F2-Score (\%)} \\ \toprule
BSP \cite{jin2011towards} & 75.12 & 72.54 & 73.80 & 73.04   \\
Devmane and Rana \cite{devmane2014detection}  & 66.32 & 68.86  & 67.56 & 68.33  \\
Goga et al. \cite{goga2015doppelganger} & 63.54 & 79.61 & 70.67& 75.77  \\ 
Kamhoua et al. \cite{kamhoua2017preventing}  & 67.41 & 71.44  & 69.36& 70.59 \\
Zheng et al. \cite{zheng2015detecting} & 68.15  & 73.34   & 70.64 & 72.23 \\
NPS-AntiClone    &  \textbf{88.70} &	\textbf{82.83} &	\textbf{85.66}& \textbf{83.94}
 \\ 
\toprule
\end{tabular}%
}
\end{table}

\subsubsection{Impact of the multi-view representation}
\begin{figure}[]
\centering
  
  \begin{tikzpicture}[thick,scale=1, every node/.style={scale=1}]
  \definecolor{bulgarianrose}{rgb}{0.28, 0.02, 0.03}
  \definecolor{clr1}{RGB}{81,82,83}
    \definecolor{clr2}{RGB}{31,182,83}
    \definecolor{clr3}{RGB}{31,18,213}
    \definecolor{clr4}{RGB}{0,32,96}
    \definecolor{clr5}{RGB}{0.36, 0.54, 0.66}
    \definecolor{cadetblue}{rgb}{0.37, 0.62, 0.63}
    \definecolor{darkjunglegreen}{rgb}{0.1, 0.14, 0.13}
    \definecolor{darklava}{rgb}{0.28, 0.24, 0.2}
    \begin{axis}[
        %/pgf/number format/1000 sep={},
        width=8cm,
        height=5cm,
        %at={(0.758in,0.981in)},
        %scale only axis,
        %clip=false,
        %separate axis lines,
        %axis on top,
        %xmin=0,
        %xmax=5,
        xtick={0,1,2,3,4,5,6,7},
        %x tick style={draw=none},
        xticklabels={NPS-AntiClone$_{PA}$,NPS-AntiClone$_{posts}$,NPS-AntiClone$_{FL}$,NPS-AntiClone$_{F}$,NPS-AntiClone$_{net(F,FL)}$,NPS-AntiClone$_{net(F,FL)+{posts}} $,NPS-AntiClone$_{concat}$,NPS-AntiClone},
        x tick label style={rotate=35,anchor=east},
        ytick={45,50,55,60,65,70,75,80,85,90},
        ymin=45,
        ymax=90,
        ymajorgrids=true,
        grid style=dotted,
        ylabel={F1-Score(\%)},
        every axis plot/.append style={
          ybar,
          bar width=.4,
          bar shift=0pt,
          fill
        }
      ]
      \addplot [bulgarianrose] coordinates {(0,53.17)};
      \addplot [clr1] coordinates {(1,55.17)};
      \addplot [clr3] coordinates{(2,50.74)};
      \addplot [clr4] coordinates{(3,53.36)};
      \addplot [clr2] coordinates{(4,61.81)};
      \addplot [darkjunglegreen] coordinates{(5,71.89)};
      \addplot [darklava] coordinates{(6,64.33)};
      \addplot [cadetblue] coordinates{(7,85.66)};
    \end{axis}
  \end{tikzpicture}
    \caption{Impact of the multi-view}
  \label{multi}
\end{figure}
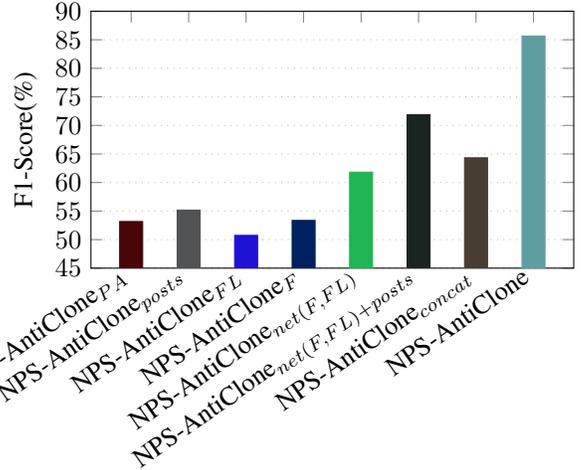

We also compared NPS-AntiClone with %its variants
the variants announced in Section IV.C, namely, NPS-AntiClone$_{posts}$, NPS-AntiClone$_{net(F,FL)}$, NPS-AntiClone$_{FL}$, NPS-AntiClone$_{F}$ and NPS-AntiClone$_{concat}$, and %show the the performance results in  
the obtained results are depicted in Figure \ref{multi}. We %see 
observed that NPS-AntiClone$_{net(F,FL)}$ performs better than the other variants, which shows the strength of the network representation when representing the account's followers and friends. NPS-AntiClone$_{net(F,FL)}$ %achieved 
achieved a F-1 score of 61.81\%, which %was 
is 11.07\% and 8.45\% higher than that of the  NPS-AntiClone$_{FL}$ and NPS-AntiClone$_{F}$, respectively. %By combining the
The approach that combines NPS-AntiClone$_{net(F,FL)}$ and NPS-AntiClone$_{posts}$, NPS-AntiClone$_{net(F,FL)+posts}$ %outperforms 
was observed to outperform all other variants by %almost 
approximately 18\% in terms of the F-1 score. Furthermore, merging the NPS-AntiClone$_{PA}$ with NPS-AntiClone$_{net(F,FL)+posts}$ improved the performance of the final proposed model by 13\% in terms of F-1 score. 

We also experimented with an alternative way to combine all the views. In that, we concatenated all the views into a single account representation. We found that NPS-AntiClone$_{concat}$ performs poorly compared to the NPS-AntiClone, and that NPS-AntiClone outperforms the NPS-AntiClone$_{concat}$ by almost 20\% in terms of the F-1 score. The performance results indicated that the %combined 
learned account multi-view using wGCCA can benefit the identity cloning detection by improving the performance.

\subsubsection{Impact of the cosine similarity score}
NPS-AntiClone depends on the cosine similarity score to decide whether one of the %account pair
accounts in a given account pair is %a 
cloned or not. %An essential question is 
Therefore, an important question this raises is how this similarity score impacts the performance of NPS-AntiClone. Figure \ref{cosine} shows the performance results of NPS-AntiClone based on %the 
different values of the cosine similarity score. We noticed that the Precision increases and Recall decrease when the similarity score %increases 
was gradually incremented from 0.1 to 0.9. When the cosine similarity score is high, it %means that less 
results in a less number of correct positive account pairs %will be 
to be detected. On the other hand, a smaller cosine similarity score %means that 
forces more negative account pairs %will 
to be detected as a pair of cloned accounts and its victim account. Furthermore, when the similarity score was increased from 0.8 to 0.9, %Precision only raises slightly yet Recall decreases significantly
although the Precision only increased slightly, Recall was observed to decrease significantly, resulting in a notable decrease in F1-Score. However,  F2-Score declined slightly when the cosine similarity %score 
score was gradually incremented from 0.1 to 0.8. Therefore, we %choose 
chose 0.1 as the optimal cosine similarity score, which leads to better results in %the 
a recall-prioritized scenario, as described before. %situation.
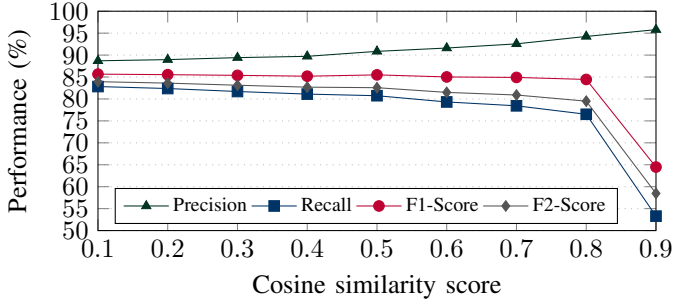
\begin{figure}[]
  \centering
  
\begin{tikzpicture} [thick,scale=1, every node/.style={scale=1}]
\definecolor{crimsonglory}{rgb}{0.75, 0.0, 0.2}
\definecolor{darkgreen}{rgb}{0.0, 0.2, 0.13}
\definecolor{darkmidnightblue}{rgb}{0.0, 0.2, 0.4}
\definecolor{britishracinggreen}{rgb}{0.0, 0.26, 0.15}
\definecolor{davy}{rgb}{0.33, 0.33, 0.33}
\begin{axis}[
    legend columns=-1,
    legend style={nodes={scale=0.75, transform shape}},
    width=9cm,
    height=4.5cm,
    title={},
    xlabel={Cosine similarity score},
    ylabel={Performance (\%)},
    xmin=0.1, xmax=0.9,
    ymin=50, ymax=100,
    xtick={0.1,0.2,0.3,0.4,0.5,0.6,0.7,0.8,0.9},
    ytick={50,55,60,65,70,75,80,85,90,95,100},
    legend pos=south west,
    ymajorgrids=true,
    grid style=dotted,
]
\addplot[
    color=darkgreen,
    mark=triangle*,
    mark size=2pt
    ]
    coordinates {
    (0.1,88.7)(0.2,88.97)(0.3,89.42)(0.4,89.72)(0.5,90.85)(0.6,91.61)(0.7,92.56)(0.8,94.25)(0.9,95.8)
    };
    
\addplot[
    color=darkmidnightblue,
    mark=square*,
    mark size=2pt
    ]
    coordinates {
    (0.1,82.83)(0.2,82.38)(0.3,81.7)(0.4,81.11)(0.5,80.74)(0.6,79.32)(0.7,78.44)(0.8,76.51)(0.9,53.29)
    };
\addplot[
    color=crimsonglory,
    mark=*,
    mark size=2pt
    ]
    coordinates {
    (0.1,85.66)(0.2,85.54)(0.3,85.38)(0.4,85.19)(0.5,85.49)(0.6,85.02)(0.7,84.91)(0.8,84.45)(0.9,64.48)
    };
\addplot[
    color=davy,
    mark=diamond*,
    mark size=2pt
    ]
    coordinates {
    (0.1,83.94)(0.2,83.61)(0.3,83.13)(0.4,82.69)(0.5,82.57)(0.6,81.50)(0.7,80.90)(0.8,79.50)(0.9,58.47)
    };
\legend{Precision, Recall, F1-Score, F2-Score}
\end{axis}
\end{tikzpicture}
\caption{Impact of the cosine similarity score}
\label{cosine}
\end{figure}

\subsubsection{%Compared with supervised machine learning
Performance comparison with supervised machine learning models}
We compared NPS-AntiClone against %the supervised models
a comprehensive set of supervised machine learning models, as well. The comparison results among NPS-AntiClone and the other machine learning and deep learning models revealed that NPS-AntiClone significantly outperformed all the supervised models (see Table \ref{zzz1}). The supervised machine learning and deep learning models require more labelled data in order to %train the model
obtain a satisfactory performance \cite{goodfellow2016deep,bach2017learning}. However, NPS-AntiClone only calculates the cosine similarity between the account pair, which does not require any training. 
\begin{table}[]
\centering
\caption{Comparison with the baseline machine and DL models evaluated as the predictor of NPS-AntiClone. Each KPI is presented as an average over 10 iterations together with standard deviation ($\sigma$).}
\label{zzz1}
\resizebox{\columnwidth}{!}{%
\begin{tabular}{p{0.21\columnwidth}p{0.18\columnwidth}p{0.17\columnwidth}p{0.17\columnwidth}p{0.17\columnwidth}}
\toprule
\textbf{Model} &   \textbf{Precision ($\sigma$)} & \textbf{Recall ($\sigma$)} & \textbf{F1-Score ($\sigma$)} &\textbf{F2-Score (\%)} \\ \toprule
ADA   & 82.64 (1.45) & 76.64 (0.87) & 79.44 (1.19) & 77.76 (0.94)\\ 
CNN   & 84.48 (0.53) & 75.75 (1.22)  & 79.87 (1.54) & 77.34 (0.96)\\ 
DNN  & 73.08 (1.33) & 78.70 (2.10)& 75.76 (0.71)& 77.50 (1.88) \\ 
KNN  & 86.8 (1.06) & 69.22 (1.04) & 76.97 (0.85)& 72.14 (1.04)  \\ 
LR & 74.14 (1.70) & 77.13 (0.59) & 75.36 (0.96)& 76.51 (1.04)   \\
MLP  & 72.06 (1.94) & 77.76 (1.25) & 74.71 (1.76)& 76.54 (1.34)   \\ 
RF & 86.64 (1.58) & 76.03 (3.58) & 80.98 (1.27)& 77.93 (2.85) \\ 
SVM & 81.37 (2.01)  & 76.13 (0.92) & 78.66 (1.40)& 77.12 (1.03) \\ %\hline
NPS-AntiClone    &  \textbf{88.70} &	\textbf{82.83} &	\textbf{85.66}& \textbf{83.94} \\
\toprule
\end{tabular}%
}
\end{table}
\subsubsection{Impact of the weight of the wGCCA}
\begin{figure}[]
\centering
  
  \begin{tikzpicture}[thick,scale=1, every node/.style={scale=1}]
    	\definecolor{coolblack}{rgb}{0.0, 0.18, 0.39}
    \begin{axis}[
        %/pgf/number format/1000 sep={},
        width=8cm,
        height=5cm,
        %at={(0.758in,0.981in)},
        %scale only axis,
        %clip=false,
        %separate axis lines,
        %axis on top,
        %xmin=0,
        %xmax=5,
        xtick={0,1,2,3,4,5,6,7,8,9},
        %x tick style={draw=none},
        xticklabels={$_{[0.5,0.5,1,0.25]}$,$_{[1.0,1.0,1.0,1.0]}$,$_{[0.25,0.5,0.5,0.25]}$,$_{[0.5,0.5,0.5,0.5]}$,$_{[0.25,0.5,0.25,0.25]}$,$_{[1.0,1.0,0.5,0.25]}$,$_{[0.25,1.0,0.5,0.25]}$,$_{[0.25,0.25,0.25,0.25]}$,$_{[0.25,0.5,0.25,0.5]}$,$_{[0.25,1.0,1.0,0.25]}$},
        x tick label style={rotate=40,anchor=east},
        ytick={45,50,55,60,65,70,75,80,85,90},
        ymin=45,
        ymax=90,
        ymajorgrids=true,
        grid style=dotted,
        ylabel={F1-Score(\%)},
        every axis plot/.append style={
          ybar,
          bar width=.4,
          bar shift=0pt,
          fill
        }
      ]
      \addplot [coolblack] coordinates {(0,82.68)};
      \addplot [coolblack] coordinates {(1,77.80)};
      \addplot [coolblack] coordinates {(2,85.66)};
      \addplot [coolblack] coordinates {(3,76.04)};
      \addplot [coolblack] coordinates {(4,73.48)};
      \addplot [coolblack] coordinates {(5,80.13)};
      \addplot [coolblack] coordinates {(6,77.19)};
      \addplot [coolblack] coordinates {(7,76.04)};
      \addplot [coolblack] coordinates {(8,72.8)};
      \addplot [coolblack] coordinates {(9,83.47)};

    \end{axis}
  \end{tikzpicture}
    \caption{Impact of the weight of the wGCCA}
  \label{weight}
\end{figure}
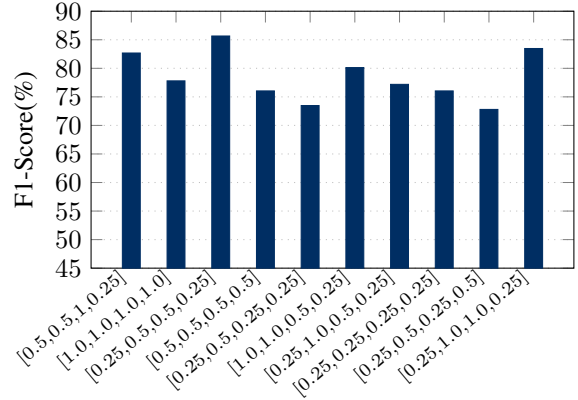

We also performed multiple rounds of experiments to study the impact of the weight of the wGCCA on each view. To do that, we tested different combinations of weight (i.e. 0.25, 0.5 and 1) for each view. We assigned a weight for each view $[posts, net_F, net_{FL}, PA]$. Figure \ref{weight} shows the top 10 results %of 
observed against different combinations of the weights. It can be seen that when the profile attribute view %are 
was given a low weight, the F1-Score %is 
was increasing. We also noticed that when the $net_F$, $net_FL$ %are 
were getting a high weight, the F1-Score %is 
was increasing. The weight of the ${posts}$ %seems that does not 
was not observed to have a high impact on the wGCCA. The optimal %weight was 
vector of weights observed for the features $[posts, net_F, net_{FL}, PA]$ was $[0.25,0.5,0.5,0.25]$.

\section{Conclusion and Future Work}
\label{rw3}
 
We proposed a novel unsupervised SocSen service provider identity cloning detection approach, NPS-AntiClone, based on non-privacy-sensitive user features. We devised a %multi-view account generation 
multi-view account representation model that generates different views for each account in an account pair %comprised of 
comprising three  categories of views, namely, 1) post view, 2) network view and 3) profile attribute view. %We then combine the generated views into a single view using wGCCA. %We finally 
We then adopted wGCCA to learn a single %representation
embedding from the generated multi-view.
Finally, we calculate the cosine similarity between the account pair. NPS-AntiClone was evaluated on a real-world Twitter dataset against other state-of-the-art cloned identity detection techniques and supervised machine learning models. The results showed that the proposed approach significantly outperformed the other models. 

In future, we plan to design a method that is able to point out which %one of the account pair 
account of an account pair is the cloned identity, since our proposed approach only detects whether an account pair form a pair of cloned accounts and its victim.
%In future, since our proposed approach only detects whether an account pair form a pair of cloned account and its victim we plan to design a method that able to point out which the cloned identity. %we plan to explore social graph-based identity cloning detection techniques and the application of our approach in other social media platforms.

\bibliographystyle{ieeetr}
\bibliography{ref.bib}

\end{document}